\documentclass[3p,times]{elsarticle}

\usepackage{ecrc}

\usepackage{graphicx}

\volume{00}

\firstpage{1}

\journalname{Nuclear Physics A}

\runauth{Heng-Tong Ding}

\jid{npa}

\jnltitlelogo{Nuclear Physics A}

\usepackage{amssymb}

\biboptions{square,comma,numbers,sort&compress}

\usepackage[figuresright]{rotating}
\usepackage{graphics}

\newcommand{\nc}[1]{\newcommand{#1}}
\nc{\beqa}{\begin{eqnarray}}
\nc{\eeqa}{\end{eqnarray}}

\nc{\beq}{\begin{equation}}
\nc{\eeq}{\end{equation}}
\nc{\hmu}{\hat{\mu}}
\nc{\nn}{\nonumber}

\nc{\vecp}{\mathbf{p}}
\nc{\md}{\mathrm{d}}
\nc{\vecx}{\mathbf {x}}
\nc{\vecnull}{\mathbf {0}}

\begin{document}

\begin{frontmatter}



\dochead{}

\title{Hard and thermal probes of QGP from the perspective of \\
Lattice QCD}


\author{Heng-Tong Ding}

\address{Key Laboratory of Quark \& Lepton Physics (MOE) and Institute of
Particle Physics, \\
Central China Normal University, Wuhan 430079, China}

\begin{abstract}
In this talk I review the current status of lattice QCD results on 
the hard and thermal probes of QGP, including jet quenching parameters, 
the melting of quarkonia and open heavy flavours, thermal photon/dilepton rates, electrical conductivity as
well as heavy quark diffusion coefficients.
\end{abstract}


\end{frontmatter}


\section{Introduction}
\label{sec:intro}

The primary goal of heavy ion programs at the RHIC and LHC experiments is
to explore the properties of nuclear matter under extreme conditions.  It
has been confirmed by experimentalists at RHIC and LHC that a new matter, i.e.
Quark Gluon Plasma (QGP) is formed in the relativistic heavy ion collisions.
However, properties of QGP are not yet fully understood quantitatively. 
This requires efforts from both the analysis of experimental data and theoretical modelling.  
Lattice QCD calculations, which are based on first principles, can provide crucial inputs to phenomenological 
studies of QGP properties. In the following I will  review the current 
status of lattice QCD calculations on the deconfinement, electromagnetic and transport properties of
QGP.

\section{Deconfinement properties of QGP}

It is now well-known that the transition of nuclear matter from hadronic phase to 
quark gluon plasma phase is not a true phase transition but a rapid crossover. The transition temperature $T_c$ of 
such a rapid crossover is around 154 MeV~\cite{hotQCD_Tc,WB_Tc}. Unlike light hadrons which probe the chiral aspect of 
the QCD transition, e.g. the degeneracy of vector meson $\rho$ and axial-vector meson $a_1$ signals the restoration of chiral symmetry, 
the fate of heavy hadrons in the hot and dense medium reflects more on the deconfinement properties of QGP, e.g. they may survive at temperatures
above $T_c$ due to their small sizes and large binding energies~\cite{HQ_seed}.

The abundance production of strange hadrons in heavy ion collisions compared to 
proton-proton collisions is considered as one of the signatures that QGP is formed. This is based on the fact that there 
are no strange valence quarks in the initial colliding nuclei. However, it is not yet clear from the theoretical side whether strange hadrons survive above $T_c$ or not. 
One may think that since the strange quark is much heavier than the light quark and is not much affected by the chiral symmetry then the strange hadrons 
may survive above $T_c$. The understanding of the fate of strange hadrons in the hot medium also affects our thinking on its freeze-out temperature.

One way to investigate the fate of open strange hadron
in the hot medium is to check the fluctuations and correlations of conserved quantum numbers, e.g. the baryon number (B)
and electric charge (Q). By comparing these quantities calculated from lattice QCD with those calculated from Hadron Resonance Gas
(HRG) model at lower temperatures one can get an idea about at what temperature the new degrees of freedom start to emerge. 

The diagonal cumulants and their off-diagonal correlations are defined as the derivatives of pressure with respect to various chemical potentials $\hat{\mu}_X=\mu_X/T$
\beq
\chi_{\rm mn}^{\rm XY} =\frac{\partial^{(m+n)} \big{(}p(\hat{\mu}_X,\hat{\mu}_Y)/T^4\big)}{\partial \hat{\mu}_X^m \partial \hat{\mu}_Y^n}\Big{|}_{\vec{\mu}=0} ,
\eeq
where $X,Y=B,Q,S,C$ and $\vec{\mu}=(\mu_B,\mu_Q,\mu_S,\mu_C)$ and $\chi_{0n}^{XY}\equiv\chi_n^Y$ and $\chi_{m0}^{XY}\equiv\chi_m^X$. 
In an uncorrelated gas of hadrons, i.e. described by the HRG model, the pressure of all the strange hadrons can be decomposed
into mesonic part $P^{HRG}_M$ and baryonic part $P^{HRG}_B$~\cite{strangeness}
\beq
P^{HRG}_S(\mu_B,\hmu_S) = P^{HRG}_{|S|=1,M} \cosh(\hmu_S)  + \sum_{\ell=1}^{3}P^{HRG}_{|S|=i,B} \cosh(\hmu_B-\ell\hmu_S).
\label{eq:P-HRG}
\eeq
Note that Boltzmann statistics is assumed here. For strange hadrons this is ensured since even the mass of the lightest strange hadron, i.e. Kaon, has the value of $\sim 3T_c$.

\begin{figure}[!th]
\begin{center} 
\includegraphics[width=0.4\textwidth]{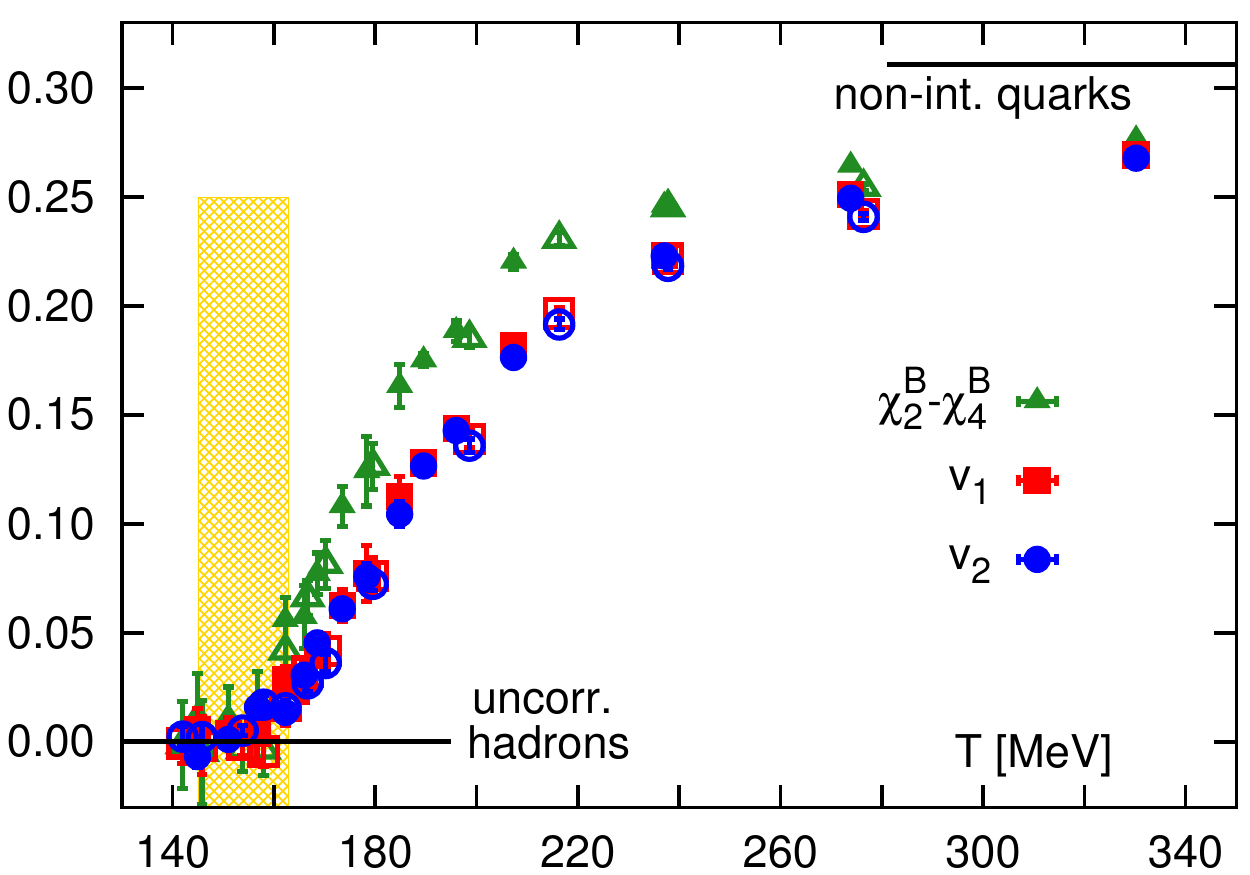}~~\includegraphics[width=0.4\textwidth]{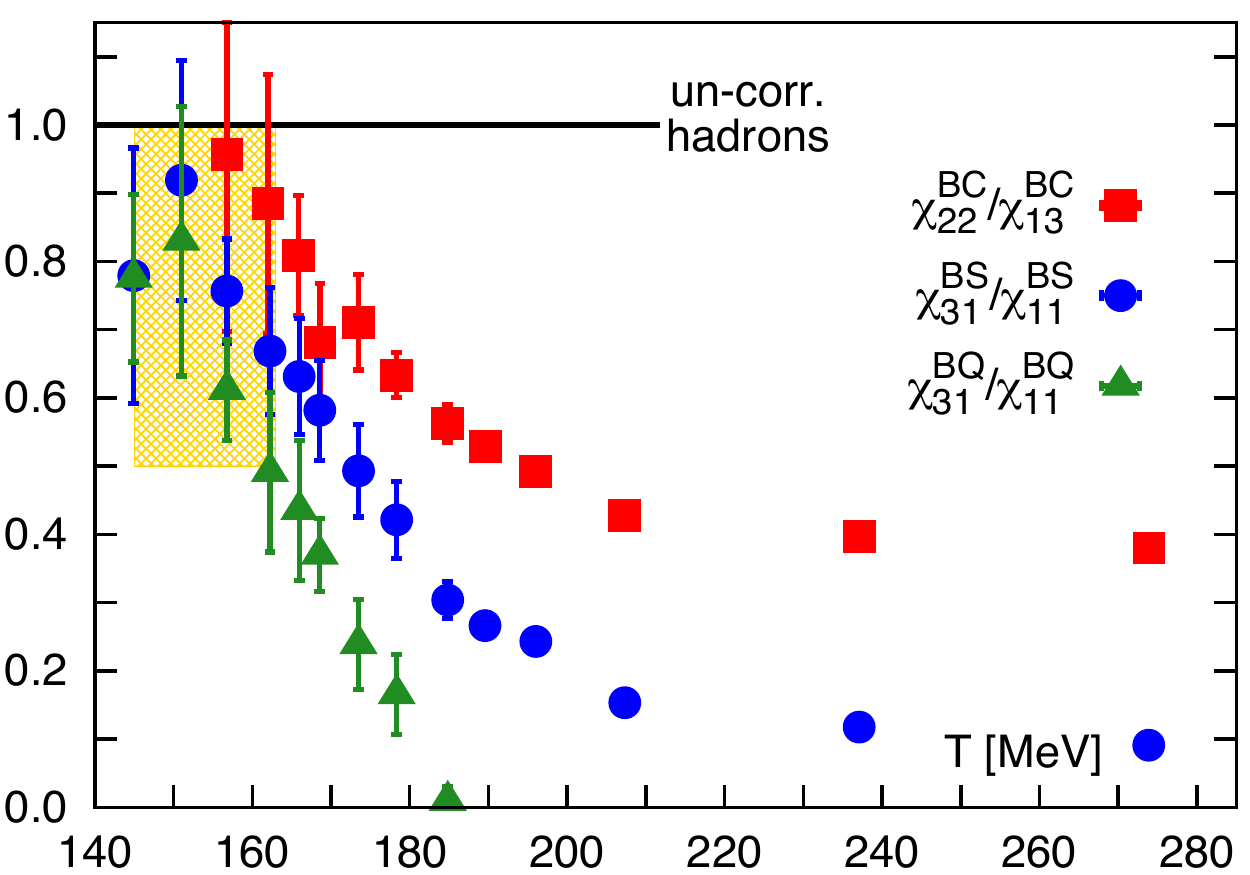} 	
\caption{Breaking-down of uncorrelated hadron resonance gas description for charge correlations involving light, strange and charm quarks at temperatures in or above
chiral crossover region as indicated by the yellow band. The left plot shows the charge correlations for light and strange sectors~\cite{strangeness} while the right plot shows the results involving light, strange 
and charm quark sectors~\cite{charmness}.}
\label{fig:fluctuations}
\end{center}
\end{figure}

Combinations of fluctuations, i.e. $v_1 = \chi^{\rm BS}_{31} - \chi^{BS}_{11}$ and $\chi_2^B-\chi_4^B$ should vanish for an uncorrelated gas of hadrons within the classical Boltzmann approximation
since the baryon-strangeness correlation differing by even number of $\mu_B$ derivatives are identical; $v_2=\frac{1}{3}(\chi_2^S-\chi_4^S)-2\chi_{13}^{BS}-4\chi_{22}^{BS}-2\chi_{31}^{BS}$
denoting the differences of two operators each corresponding to the partial pressure of all strange hadrons, should also vanish in the uncorrelated hadron resonance gas. 
Here $v_1$ and $v_2$ receive contributions only from strange hadrons while $\chi_2^B-\chi_4^B$ receive contributions dominantly from light-quark hadrons. It is apparent from the left plot of Fig.~\ref{fig:fluctuations} 
that all these three quantities deviate from the value of zero given by HRG almost at the same temperature, i.e. in the vicinity of
chiral crossover temperature as indicated by the yellow band. This suggests that the strange degrees of freedom are liberated from hadrons at almost same temperature as light quarks.
The whole analysis has been repeated using a different fermion action in Ref.~\cite{Bellwied:2013cta}.

Based on the same idea one can also discuss the fate of charmed hadrons in the hot medium~\cite{charmness}. As in an uncorrelated hadron resonance gas the charmed baryon sector with
charm quantum number $C=\pm 1$ dominates the contribution to the partial pressure of all charmed hadrons, $\chi_{nm}^{BC}\simeq\chi_{11}^{BC}$
with $n+m$ even and larger than 2. Thus three quantities, $\chi_{22}^{BC}/\chi_{13}^{BC}$, $\chi_{31}^{BS}/\chi_{11}^{BS}$ and $\chi_{31}^{BQ}/\chi_{11}^{BQ}$, which receive
contributions only from charm, strange and light quark sectors, respectively, should be equal to unity within a description by an uncorrelated hadron resonance gas model. As can be clearly seen  from the right plot of Fig.~\ref{fig:fluctuations} 
that such a description breaks down for baryonic correlations involving light, strange, or charm quarks in or just above
the chiral crossover region.

The above study through the fluctuations of conserved quantum numbers give the information on the properties of open heavy hadrons at finite temperature.
However, it is not able to tell anything on the fate of open heavy hadrons in specific channels and bound states with zero net charm and strange quantum number.
To investigate the fate of various species of bound states one has to look into their spectral functions. The spectral functions $\rho_H(\omega,T)$ are related with the two-point Euclidean 
correlation functions as follows
\beqa
\label{eq:Gtau}
G_H(\tau,T) &=& \int_0^\infty \mathrm{d}\omega \,\frac{\cosh(\omega(\tau -1/2T))}{\sinh (\omega/2T)} \, \rho_H(\omega,T), \\
G_H(z,T) &= &\int_0^\infty  \frac{2 \mathrm{d}\omega}{\omega} \int_{-\infty}^{\infty} \mathrm{d}p_z e^{ip_z z} \, \rho_H(\omega,p_z,T).
\eeqa
Here $G_H(\tau,T)$ and $G_H(z,T)$ are temporal and spatial correlation functions in a hadronic channel $H$, respectively. As shown in the above equations
the relation between spectral function and temporal correlation function is more straightforward, the spectral functions are thus commonly extracted 
from the temporal correlation functions rather than the spatial correlation functions. However, the extraction of spectral function are hampered by two issues:
one is that the maximum temporal distance is restricted by the inverse temperature and the other one is the analytic continuation. The former one basically means
that the physics in the temporal correlation functions are more compacted at higher temperatures and the latter tells that a large number of data points
in the temporal direction is crucially required. Certain techniques, e.g. Maximum Entropy Method are normally used to extract the spectral functions~\cite{mem}.

The current state of art quenched lattice QCD study on charmonium spectral functions suggests that all charmonia are already dissociated 
at 1.5 $T_c$~\cite{charm}. However, their dissociation temperatures have not yet been found. Recently an explorative lattice QCD study of charmonia properties in full QCD 
has been also made~\cite{charm2}. When moving to the case of heavier bound states, i.e. bottomonia, one can also perform lattice study based on an effective theory, i.e. Non-Relativistic QCD
due to the large mass of bottom quark and the existence of certain hierarchies in the system. One pitfall of such study is that continuum limit is not available 
by definition, while the advantage is that the relation between temporal correlators and spectral functions becomes less complicated as the thermal boundary is
removed. Similar results have been shown in recent studies on the fate of S-wave and P-wave bottomonia at finite temperature in both 2-flavour as well 
as 2+1-flavour QCD, i.e.  S wave ground states survive up to at least 2 $T_c$ while P wave ground states melt just above $T_c$~\cite{Bottomonia}. 

\begin{figure}[!th]
\begin{center} 
\includegraphics[width=0.32\textwidth]{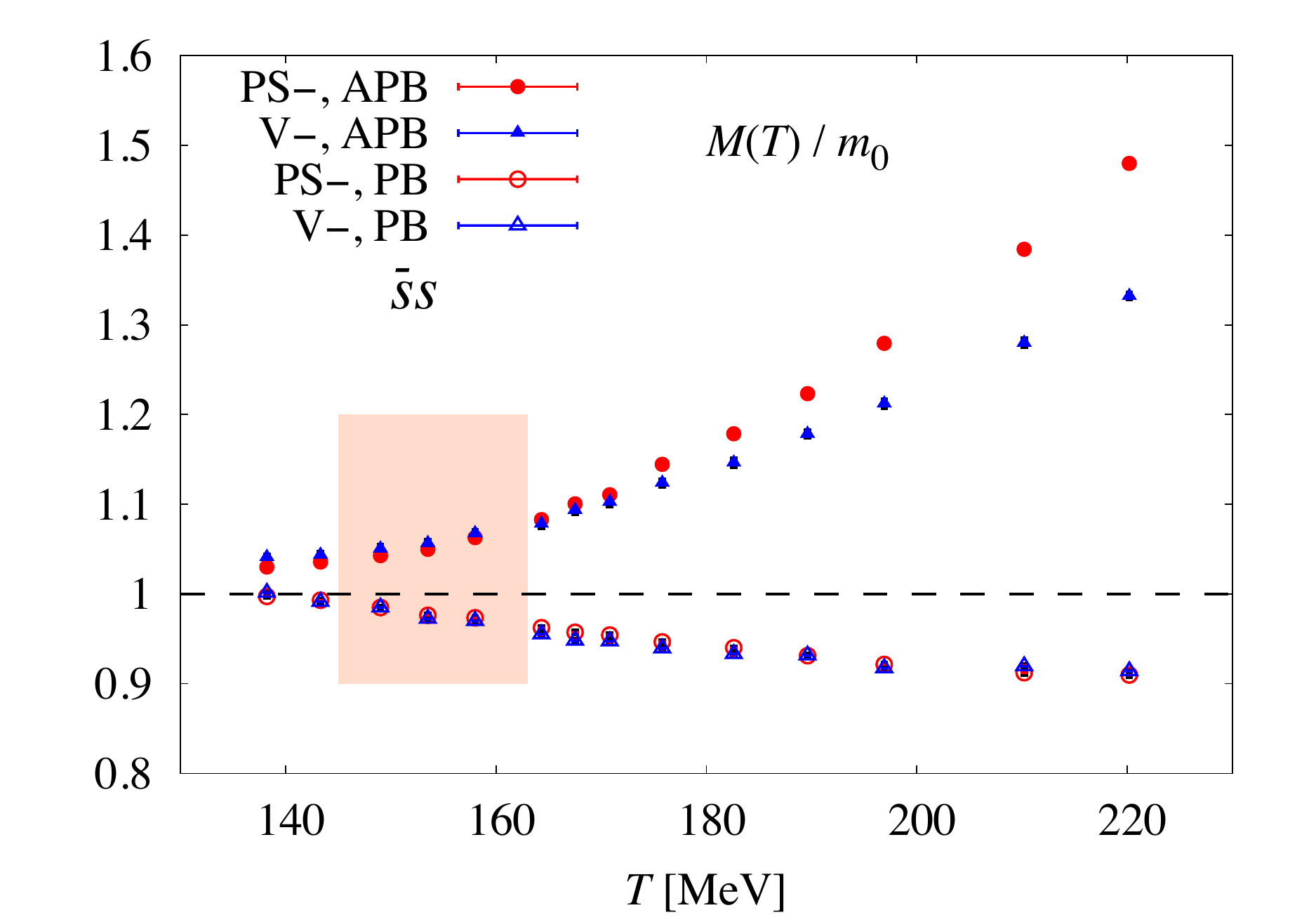} ~\includegraphics[width=0.32\textwidth]{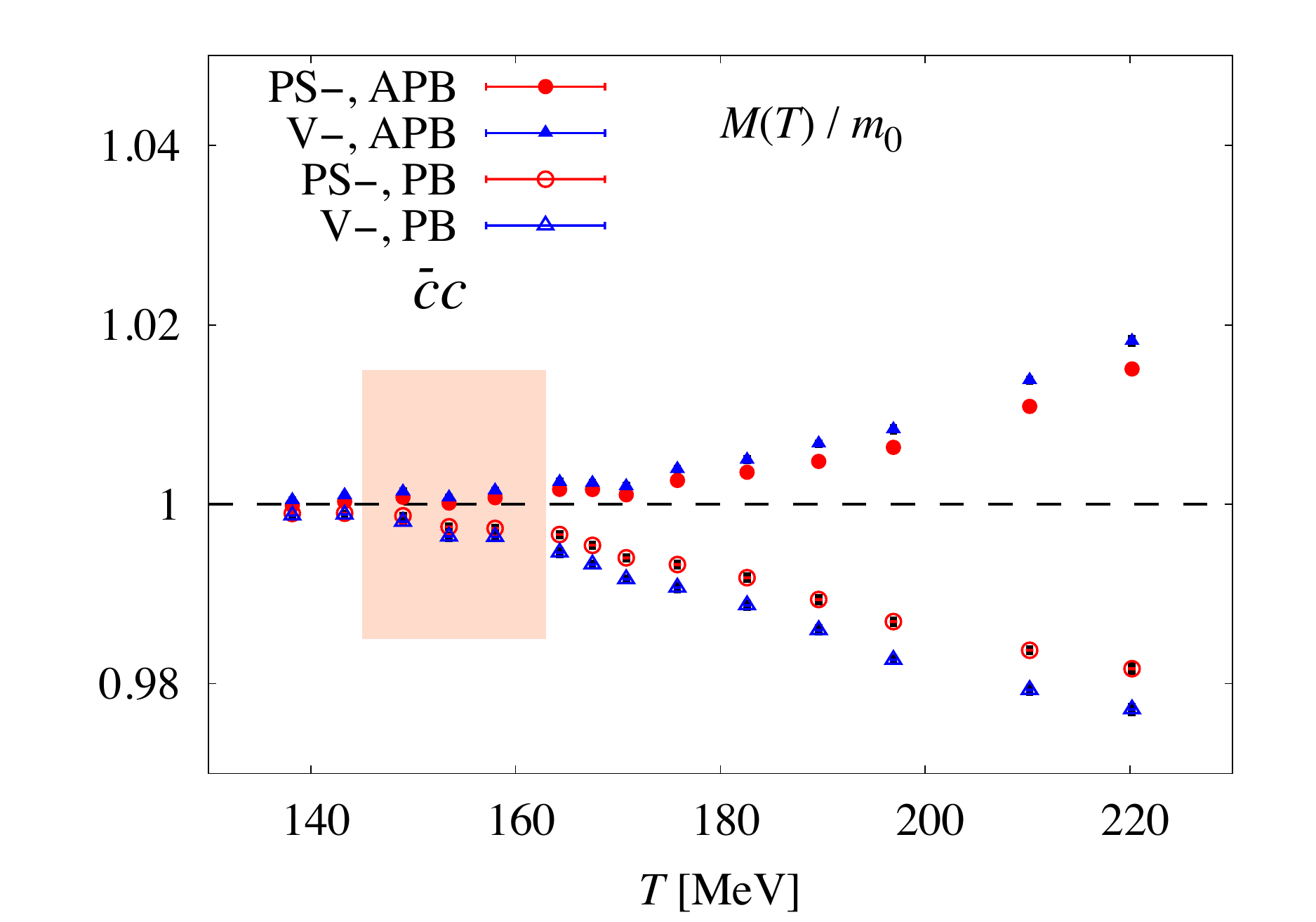}
~~\includegraphics[width=0.32\textwidth]{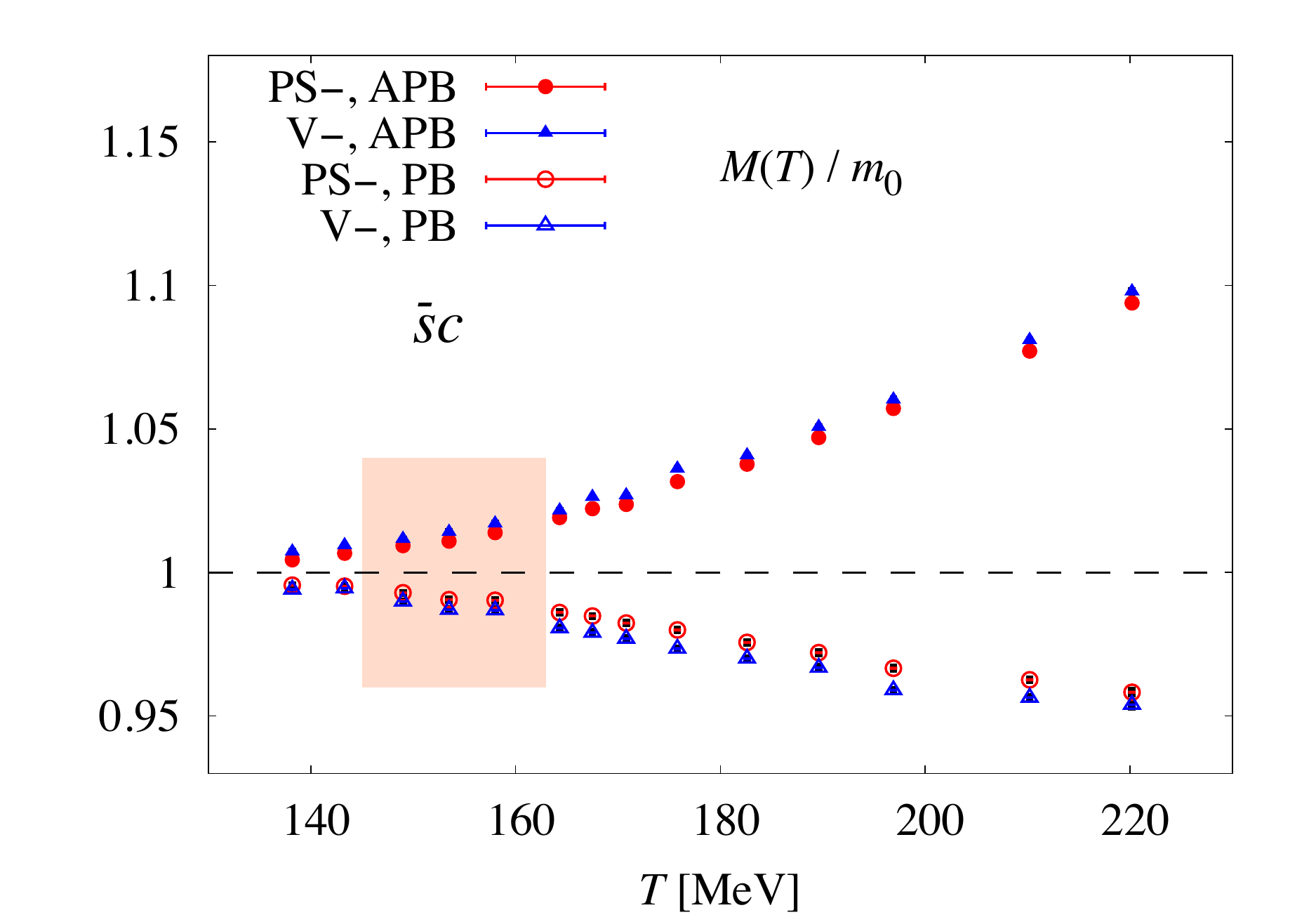} 
\caption{Ratios of screening masses at finite temperature to those at zero temperature for $s\bar{s}$ (left), $c\bar{c}$ (middle) and $\bar{s}c$ (right) systems. "PS-" stands for pseudo scalar channel while "V-" stands for vector channel.
"APB" denotes anitiperiodic boundary condition and "PB" stands for periodic condition. Figures are taken from Ref.~\cite{Maezawa:2013nxa}.}
\end{center}
\label{fig:scr}
\end{figure}

The modification of spectral function also manifests itself in the spatial correlation functions. As the physical extent in the spatial 
direction is not restricted by the temperature one can extract the exponentially decay constant, i.e. screening mass from 
the long distance behaviour of spatial correlators. At zero temperature a peak of low lying bound state with mass $m_0$ can be 
described by a spectral function $\rho(\omega,p_z,T)\sim\delta(\omega^2-p_z^2-m_0^2)$, thus the screening mass $M$ is the same 
as the pole mass $m_0$. In the high temperature limit a quark and its anti-quark are not bound anymore and the screening mass of the free quark
pair is $2\sqrt{(\pi T)^2 + m_q^2}$, where $m_q$ is the quark mass and $\pi T$ is the lowest fermionic Matsubara frequency. The appearance 
of $\pi T$ in the free limit is due to the antiperiodicity (APB) of fermions along the time direction. Thus screening mass of a mesonic bound state 
is expected to be not sensitive in the boundary conditions, i.e. with periodic boundary conditions at T=0 screening mass is still $m_0$ while in the free limit 
it becomes $2m_q$. Changing the boundary conditions and comparing with the two temperature limit one can gain some information on the modification of bound states~\cite{Karsch:2012na}.

Full lattice QCD results on the ratio of screening mass to the pole mass for $s\bar{s}$, $c\bar{s}$ and $c\bar{c}$ states in both pseudo scalar and vector channels are shown in Fig.~\ref{fig:scr}~\cite{Maezawa:2013nxa}.
It is apparent from the left plot of Fig.~\ref{fig:scr} that the bound states of $s\bar{s}$ suffer significant modifications already in the chiral crossover temperature region. 
This is consistent with the picture of unbounded open strange states in or above $T_c$ obtained from the study on the fluctuations of conserved quantum numbers
seen in Fig.~\ref{fig:fluctuations}. For the $c\bar{c}$ states, only $\sim$2\% deviation of $M$ from $m_0$ arises at 220 MeV, i.e. $\sim 1.4 T_c$. This plausible inconsistency with
the quenched lattice QCD study on temporal and spatial correlation functions~\cite{charm,Ohno:2013rka} seems to stem from the effects of dynamic quarks, e.g. in the quenched limit the phase transition is first order while
in the full QCD it is a rapid crossover. In the right plot of Fig.~\ref{fig:scr} results for open charm states $s\bar{c}$ are shown. The magnitude of deviation is larger than that of
$c\bar{c}$ states. This further strengthen the point that open charm hadrons starts to melt at almost same temperature as light and open strange hadrons as seen from Fig.~\ref{fig:fluctuations}.

Another recently developed approach to study the properties of heavy quarkonia states is based on the heavy quark effective theory.
One of the main outcomes is that the static heavy quark potential has both imaginary and real parts~\cite{Laine:2006ns}. These parts can be obtained through lattice
QCD calculations on the correlators of Wilson lines. Serval studies have been performed along this interesting line~\cite{Burnier:2013nla,Bazavov:2014kva}
and this topic was covered in much more details by A. M\'ocsy in this conference~\cite{Mocsy}.

\section{Electromagnetic properties of QGP}
\label{sec:intro}

Dileptons/photons are good probes of QGP properties as they do not participate in the strong interaction and
can carry the information of QGP produced in the early stage. However, they are produced almost during all the stages of heavy ion collisions.
It is thus important to have a detailed knowledge of dilepton/photon rates radiated from QGP. The thermal dilepton/photon rates are related to vector spectral functions in
the following way:
\beqa
\frac{\md W_{l^{+}l^{-}}}{\md\omega\,\md^3\vecp} &=& C_{em}\frac{\alpha_{em}^2}{6\pi^3}\frac{1}{\omega^2(e^{\omega/T} -1)}\,\rho_V(\omega,\vecp,T),\\
\omega \frac{{\rm d} R_\gamma}{{\rm d}^3p} &=&C_{em}\, \frac{\alpha_{em}}{4\pi^2} \,
\frac{\rho_{V}(\omega =|\vecp|, T)}{{\rm e}^{\omega/T} -1} \ .
\eeqa
where $\alpha_{em}$ is the electromagnetic fine structure constant and $C_{em}$ is the sum of the square of the elementary charges of the quark flavor $f$, $C_{em}=\sum_f Q_f^2$.
The emission rate of soft photons is also related to electrical conductivity $\sigma$ as follows
\beq
\lim_{\omega \rightarrow 0} \omega \frac{{\rm d} R_\gamma}{{\rm d}^3p} =
\frac{3}{2\pi^2}\, \sigma(T) \,T \,\alpha_{em} \ ,
\label{softphoton}
\eeq
where $\sigma(T)$ is defined as the limiting behaviour of spectral function
\beq
\frac{\sigma}{T} = \frac{C_{em}}{6} \lim_{\omega \rightarrow 0} \lim_{\vecp \rightarrow 0} 
\frac{\rho_{ii}(\omega,\vecp, T)}{\omega T} \;.
\eeq

\begin{figure}[!th]
\begin{center} 
\includegraphics[width=0.32\textwidth]{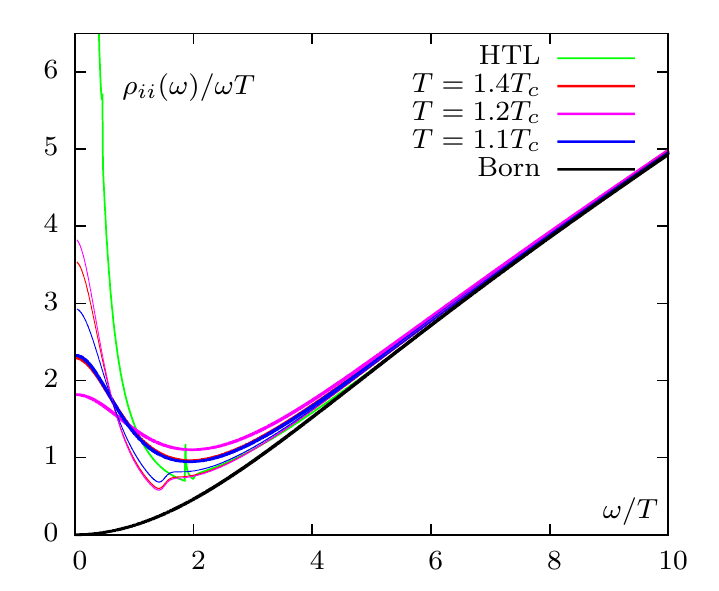}~\includegraphics[width=0.32\textwidth]{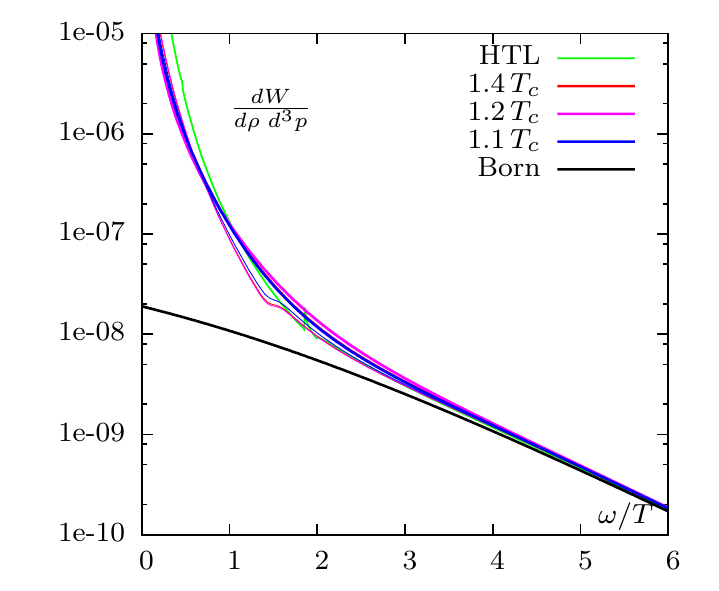}~\includegraphics[width=0.32\textwidth]{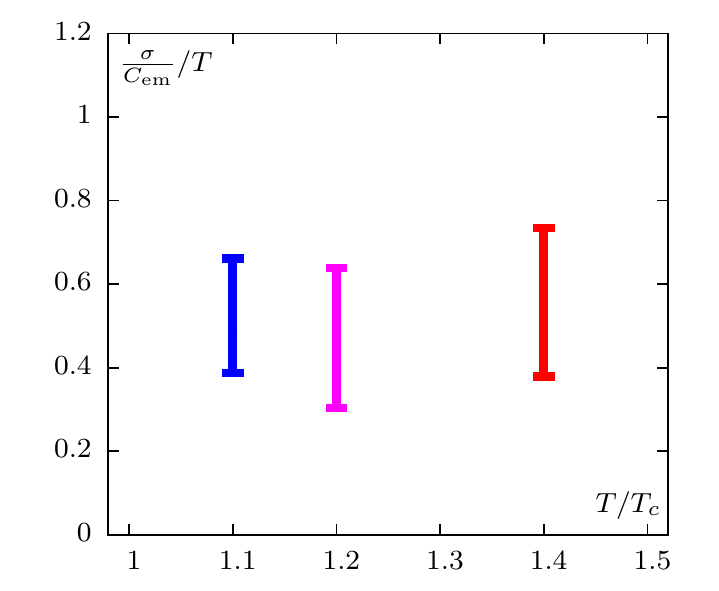}
\caption{Temperature dependences of thermal spectral functions (left), thermal dilepton rates (middle) and electrical conductivities (right) at three different temperatures above $T_c$ obtained from
quenched lattice QCD calculations~\cite{Kaczmarek:2013dya}.}
\end{center}
\label{fig:dilepton}
\end{figure}

Continuum extrapolated quenched lattice QCD results on thermal spectral functions and their resulting thermal dilepton rates as well as electrical conductivity
are shown in Fig.~\ref{fig:dilepton}.
In the high frequency region, i.e. $\omega/T \gtrsim 7$, all the thermal spectral functions at 1.1, 1.2 and 1.4 $T_c$ can be well
described by the Born rate as well as the Hard Thermal Loop (HTL) calculations. In the lower frequency region it is obvious that the lattice QCD
results differ from the Born rate and HTL calculations. The negligible temperature dependence of thermal dilepton rate and $\sigma(T)/T$ as shown in
the middle and right plots of Fig.~\ref{fig:dilepton} may due to
the fact that the phase transition in the pure glue system is just a first order and  at  $T>T_c$ the temperature is thus the only relevant scale
in the system.

\begin{figure}[!th]
\begin{center} 
\includegraphics[width=0.4\textwidth]{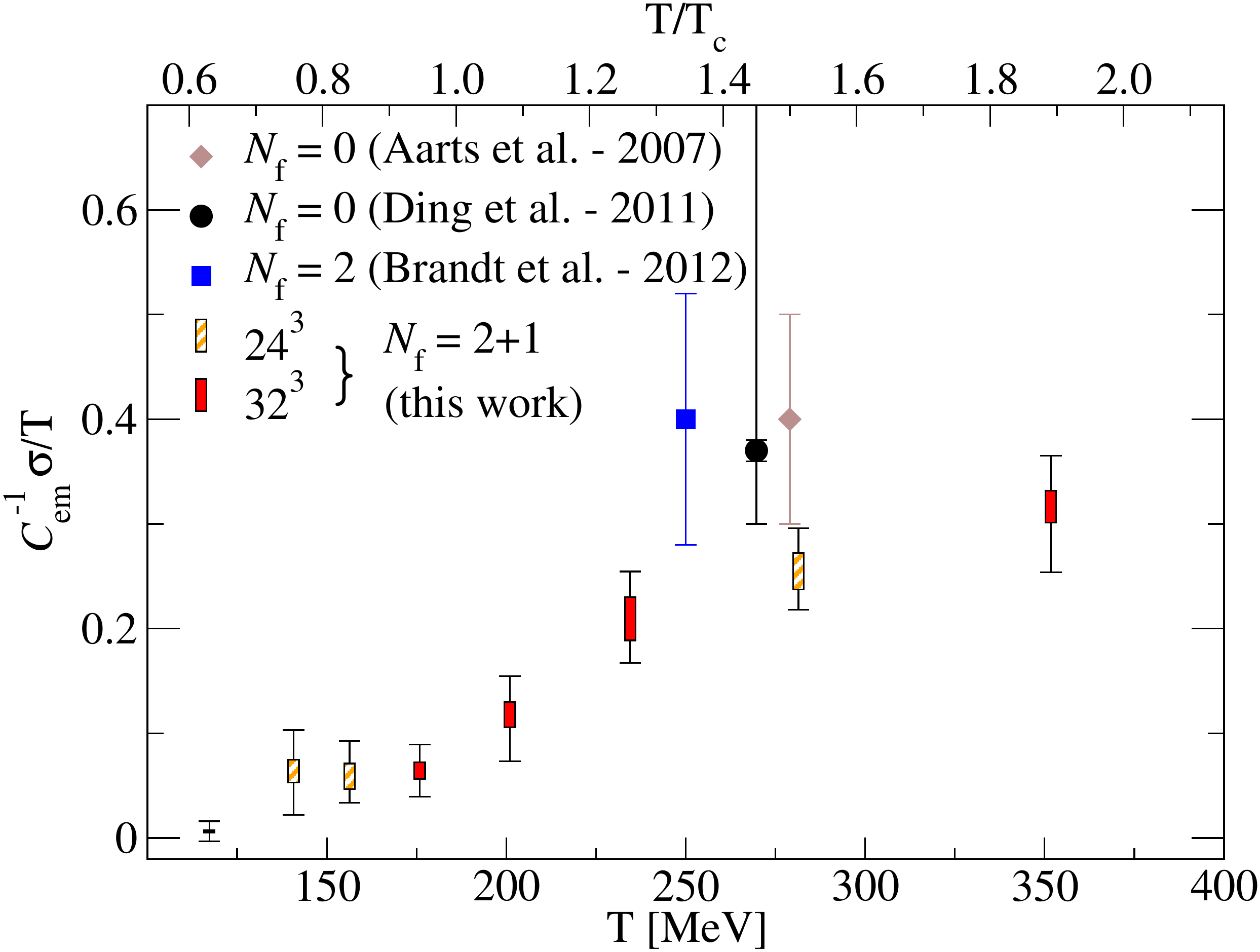}~\includegraphics[width=0.4\textwidth]{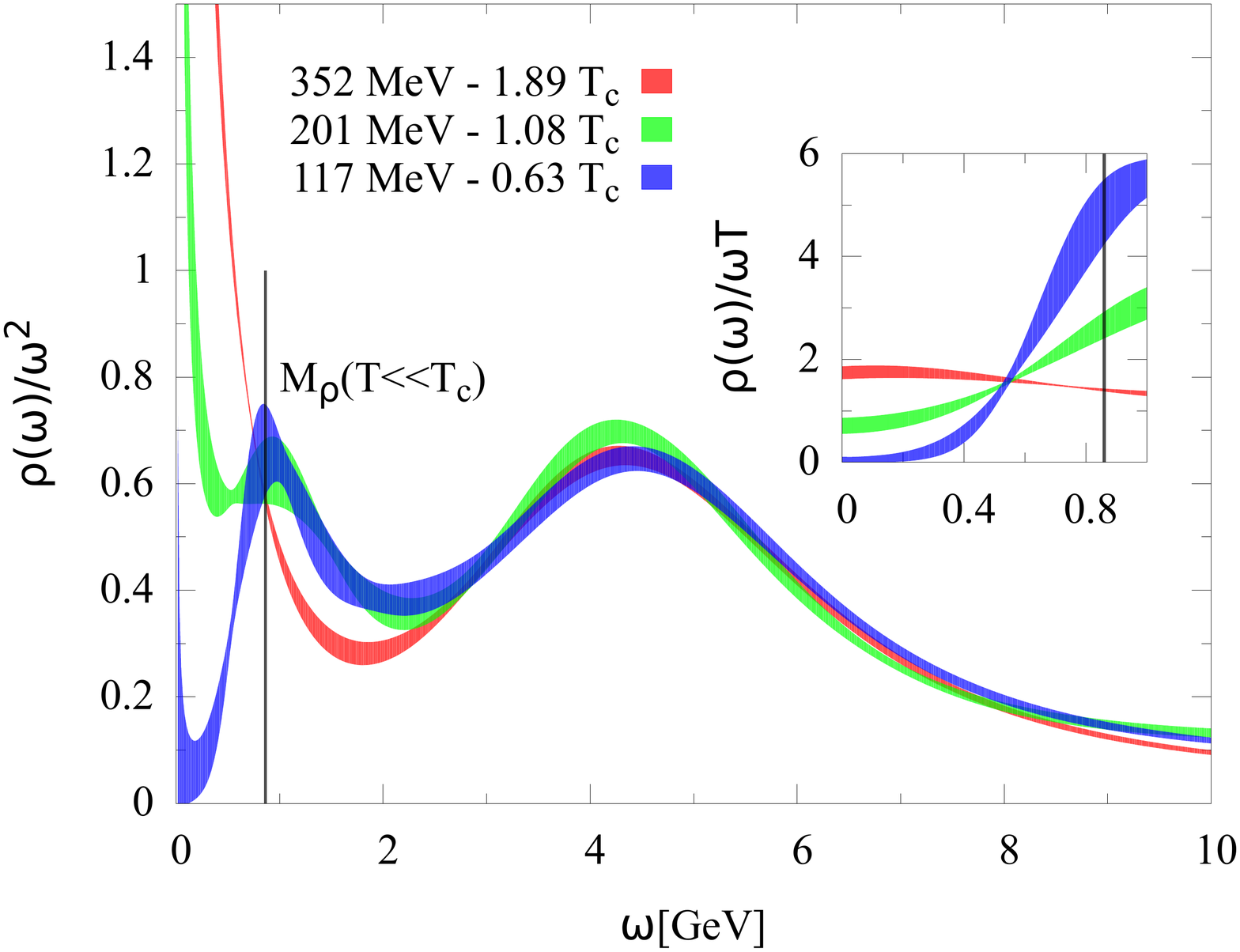}
\caption{Temperature dependences of electrical conductivities (left) and spectral functions at three temperatures (right). Figures are taken from Ref.~\cite{SwanseaSPF}.}
\label{fig:cond}
\end{center}
\end{figure}

The lattice QCD studies on electrical conductivities have been extended to full QCD recently~\cite{SwanseaSPF}.  The temperature dependence of electrical conductivity from recent
calculations in both systems with and without dynamics quarks are summarised in the left plot of Fig.~\ref{fig:cond}~\cite{dilepton,MainzSPF,SwanseaSPF}.
At $T>T_c$ there is an obvious increasing of $\sigma(T)/T$ with temperature. This is different from the case in the pure glue system as shown in the right plot of 
Fig.~\ref{fig:dilepton}. Below $T_c$ $\sigma(T)/T$ seems to first remain as constant and then drop down to nearly zero at around $0.6~T_c$. 
The observed vanishing of electrical conductivity at $0.6~T_c$ might be due to the insensitivity of the temporal correlators to the low frequency behaviour of the spectral function.
Besides the peak of $\rho$ meson with unphysical width at 0.63 $T_c$ as seen from the right plot of Fig.~\ref{fig:cond} may also make the extraction of
electrical conductivity more unreliable.
Nevertheless it might be interesting to have a look at the so-called charge diffusion coefficient, i.e. $\sigma$ divided by light quark number susceptibilities, and have a comparison with the 
heavy quark diffusion coefficient as they probably have similar temperature dependence. However, to get a quantitative determination of electrical conductivities of a QCD system with 
dynamical quarks further lattice studies are still needed. This is due to fact that the main difficulty in practice to extract spectral functions from temporal correlators is the number of points in the temporal direction ($N_\tau$).
And obviously there is no reason that uncertainties of results obtained from temporal correlators with smaller $N_\tau$, e.g. in Ref.~\cite{SwanseaSPF},
should be smaller than those shown in Ref.~\cite{dilepton}. The systematic uncertainties from analytic continuation also need to be examined using 
different methods\cite{Burnier:2012ts}.

\section{Transport properties of QGP}
\label{sec:intro}

The Brownian motion of a heavy quark in the hot medium can be described by Langevin equation whose
crucial input parameter is the heavy quark diffusion coefficient $D$. Based on the linear response theory, this dynamical
quantity can be expressed as the low frequency limit of the vector spectral function obtained from Euclidean two-point correlation functions.
The temperature dependence of charm quark diffusion coefficient $D_c$ has been studied on the quenched lattice.
It is obtained that $D_c$ is compatible with zero at $T<T_c$ and $2\pi T D_c$ is independent of temperature and is around 2 in the temperature region
from $\sim1.5~T_c$ to $\sim3~T_c$~\cite{charm}. These results are much smaller than those obtained from pQCD calculations~\cite{Diffusion_pQCD}.

Instead of extracting heavy quark diffusion coefficient from a meson spectral function as in Ref.~\cite{charm}, one can make use of Heavy Quark Effective Theory to define 
the diffusion coefficient through the low frequency limit of the spectral function related with a certain Euclidean correlation function involving color-electric fields along a Polyakov loop~\cite{CaronHuot:2009uh}. 
The usual inversion problem to obtain spectral functions from correlators still exists in this approach, however, the advantage is that the spectral function 
at low frequency  is expected to be flat and does not receive contributions from hadron states appearing in meson spectral functions. Consistent estimates on $DT$ have
been obtained by two different groups~\cite{HQEFT_diffusion1,HQEFT_diffusion2} and these results are 2$\sim$3 times larger the charm quark diffusion coefficient obtained in Ref.~\cite{charm}. 
Systematic uncertainties as well as continuum limit are certainly needed in the further study on diffusion coefficients in both studies based on QCD and HQEFT.

Jet quenching parameter $\hat{q}$, which is generally used to quantify the energy loss and multiple scatterings of energetic partons 
traversing in the medium, is another important transport coefficient of the medium. The difficulty to evaluate this quantity in lattice QCD
is simply that $\hat{q}$ is defined along the light cone with real Minkowski time $t$. Several proposals have been made recently to 
calculate $\hat{q}$ on the lattice~\cite{Ji:2013dva,Majumder:2012sh,CaronHuot:2008ni}. Among them the attempt to evaluate $\hat{q}$ on the lattice 
using a purely Euclidean and dimensionally reduced effective theory, i.e. electrostatic QCD (EQCD), is based on the
fact that the differential collision rate of partons with the medium from soft modes is almost time independent~\cite{CaronHuot:2008ni}.
Preliminary results at two temperatures, i.e. $\sim 2T_c$ and  $\sim10 T_c$ have been obtained~\cite{Panero:2013pla}. At these temperatures contributions from ultrasoft modes might be small
as results from EQCD can describe the lattice data of second and fourth order of quark number 
 susceptibilities reasonably well at temperatures above 2$T_c$~\cite{highT_qns}. However, further studies, e.g. with better improved action and finer lattice spacings
 are still needed.

%
\bibliographystyle{elsarticle-num}
\bibliography{<your-bib-database>}



\end{document}